# Single plasmon generation in an InAs/GaAs quantum dot in a transfer-printed plasmonic microring resonator

Akihito Tamada*[†], Yasutomo Ota[‡], Kazuhiro Kuruma[†], Katsuyuki Watanabe[‡], Satoshi Iwamoto[†, ‡], Yasuhiko Arakawa[‡]

[†] Institute of Industrial Science, University of Tokyo, Meguro, Tokyo 153-8505, Japan
[‡] Institute for Nano Quantum Information Electronics, Meguro, Tokyo 153-8505, Japan

**ABSTRACT:** We report single plasmon generation with a self-assembled InAs/GaAs quantum dot embedded in a plasmonic microring resonator. The plasmonic cavity based on a GaAs microring is defined on an atomically-smooth silver surface. We fabricated this structure with the help of transfer printing, which enables the pick-and-place assembly of the complicated, heterogeneous three dimensional stack. We show that a high-order surface-plasmon-polariton transverse mode mediates efficient coupling between the InAs/GaAs quantum dots and the plasmonic cavity, paving the way for developing plasmonic quantum light sources based on the state-of-the-art solid-state quantum emitters. Experimentally, we observed Purcell-enhanced radiation from the quantum dot coupled to the plasmonic mode. We also observed a strong anti-bunching in the intensity correlation histogram measured for scattered photons from the plasmonic resonator, indicating single plasmon generation in the resonator. Our results will be important in the development of quantum plasmonic circuits integrating high-performance single plasmon generators.

Plasmonic structures are a powerful tool for downscaling and highly-functionalizing integrated quantum photonic circuits owing to the tight light confinement relying on surface plasmon polariton (SPP) modes and to the resulting enhancement of light-matter interactions[1-3]. So far, various plasmonic components as building blocks for quantum circuits have been demonstrated, such as waveguides[4], beam splitters[5], single photon detectors[6] and nonlinear plasmon-plasmon gates[7]. In particular, single plasmon generators (SPGs), which deliver a stream of single SPPs[8-13], have been under intensive development. SPGs are realized by combinations of metal-based plasmonic structures with single photon emitters, such as colloidal quantum dots (QDs)[8,14] and defects in two-dimensional materials[11,15]. Meanwhile, in order to perform scalable quantum information processing using multiple plasmons/photons, it is in general required to employ SPGs with indistinguishable outputs with near-unity efficiency. In this regard, the use of self-assembled InAs/GaAs QDs is highly promising as they have already been demonstrated to potentially fulfill the demanding requirements as solid-state single photon emitters[16-18]. Another important issue is to employ high quality plasmonic structures in order to maximize the SPG performances including single plasmon generation efficiency. It should be avoided to use rough metal surfaces for supporting SPP modes, since they scatter photons out of the SPP modes and can be a source of unwanted loss in SPGs[19-21].

Here, we demonstrate a SPG based on the combination of a self-assembled single InAs/GaAs QD and a plasmonic microring resonator defined on an atomically-flat silver surface. To fabricate such a structure, we employ transfer printing method, by which GaAs microrings containing QDs are integrated on the ultra-flat silver surface in a simple pick-and-place manner. We show that a higher-order transverse plasmonic mode supports strong electric field parallel to the sliver surface, thus enabling efficient coupling to QDs with strong optical transition dipoles normal to the growth direction. We experimentally demonstrate that the high coupling between the plasmonic cavity mode and a QD manifests itself as a large Purcell effect on the QD, which leads to direct generation of quantized SPPs in the resonator from the QD. Furthermore, we performed micro-photoluminescence (PL) measurements and observed single plasmon generation from the single QD resonant to the plasmonic cavity mode.

A schematic of the plasmonic microring resonator investigated in this study is shown in Figure 1(a). The resonator consists of a GaAs microring with a diameter, height and waveguide width of a few μm, 150 nm and a few hundreds of nm, respectively. The GaAs material embeds three layers of InAs/GaAs self-assembled QDs grown by the Stranski-Krastanov mode (areal density of ~ $3.0 \times 10^{10}$ cm$^2$ per layer). The bottom QD layer is located 15-nm above the GaAs bottom edge and the distances between the QD layers are 30 nm. The ensemble of these QDs emits light with an emission peak at a wavelength of ~1200 nm at ~10 K. The microring is placed on an atomically-flat silver film, typically supporting a few SPP transverse modes at the GaAs/silver interface. For the ring diameters under the consideration, the SPP longitudinal modes constitute several whispering galley resonance modes across QD's emission band.

We construct the metal-semiconductor hybrid structure by employing transfer printing[22–24]. As the first step of the sample fabrication, the GaAs microring structures are patterned on a GaAs slab with QDs by standard semiconductor nanofabrication processes, such as electron beam lithography and dry/wet etching. The microrings are in a form of airbridge supported by small triangular contacts, as displayed in a scanning electron microscope (SEM) image for a typical microring in Figure 1(b). In parallel, we prepare an atomically-flat silver film by depositing 80-nm-thick silver on a Si(111) substrate[25] with an electron-beam evaporator and by subsequently annealing it at 450°C in a nitrogen atmosphere[21]. The root-mean-square roughness of the silver surface is only 0.3 nm, confirmed by atomic force microscopy. The smooth metal surface is important for suppressing propagation loss in SPP modes and thus for improving the efficiency of SPGs[19,20].

Then, we pick up a microring by attaching a transparent rubber stamp made of poly-dimethyl siloxane (PDMS) on the GaAs surface and quickly peeling it off. This pick up step is illustrated in Figure 1(c). Subsequently, the lifted microring is placed onto the silver surface and the PDMS stamp is slowly peeled off, leaving the microring on the silver surface, as shown schematically in the Figure 1(d). A SEM image of a fabricated plasmonic microring resonator with a diameter of 1µm and a waveguide width of 300 nm is shown in Figure 1(e). We note that the actual transfer processes were performed using a home-made transfer apparatus equipped with an optical microscope and piezo positioners[26].

We numerically analyze the plasmonic ring resonator using electromagnetic simulations. In general, the optical transition dipoles of InAs/GaAs QDs dominantly direct perpendicular to the crystal growth direction. Accordingly, they tend to couple weakly to transverse-magnetic (TM) modes, such as plane-wave SPP modes supported in infinitely-wide metal-dielectric interfaces. Figure 2(a) shows a simulated cross-sectional electric field distribution for the fundamental plasmonic mode, which exhibits a typical TM-like mode distribution with strong electric field normal to the metal surface. Whereas, the first higher-order SPP transverse mode shown in Figure 2(b) contains a certain amount of electric field components parallel to the metal plane ($E_x$). Figure 2(c) and (d) summarize the distribution of in-plane electric filed intensity for the two transverse SPP modes. In the plots, we also took into account the electric field component directed toward the wave propagation direction ($E_y$). Apparently, the higher-order SPP transverse mode supports stronger in-plain electric field, suggesting its better coupling to the embedded QDs. We numerically calculated Purcell factors ($F_p$s) in whispering galley modes based on the two SPP transverse modes using the following simple equation:

$$F_p = \frac{3}{4\pi^2}\frac{Q}{V}\left(\frac{\lambda}{n}\right)^3 \frac{|\mathbf{d}\cdot\mathbf{E}(\mathbf{r}_{QD})|^2}{|\mathbf{d}|^2|\mathbf{E}_{max}|^2}\frac{(\omega/2Q)^2}{(\omega-\omega_c)^2+(\omega/2Q)^2}, (1)$$

where $Q$ is the quality factor of the resonant mode, $V$ is the mode volume, $d$ is the electric dipole moment of the emitter, $E(r_{QD})$ is the electric field vector of the resonant mode at the position of QDs, and $\omega$ and $\omega_c$ is the angular frequency of the emitter and the resonance mode, respectively. We assume the refractive index of GaAs ($n$) to be 3.4. Importantly, equation (1) is an approximated expression and is valid only when strong material absorption and dispersion are absent[27,28]. We consider that this condition is fulfilled by the current resonator, which indeed supports moderately-high Qs with Lorentzian-shaped spectral resonances, as we will discuss below. Using the finite difference time domain method, we simulate field distributions, cavity $Q$ factors and $V$s. In the analysis, we limit ourselves in analyzing cavity modes around a wavelength of $\lambda$ = 1070 nm. To model the silver layer, we employed its material parameters at 10 K[29]. For a typical resonator design consisting of a microring with a diameter of 1 μm and a waveguide width of 300 nm, a $Q$ factor of 570 (580) and a $V$ of 1.3×10$^{-2}$ μm$^3$ (1.6×10$^{-2}$ μm$^3$) are obtained for the cavity mode based on the fundamental (higher-order) SPP transverse mode.

Then, we calculated the maximum possible Purcell effect on a certain QD embedded within the structure. We assume that the target QD lies somewhere within the QD layers and its dipole is completely parallel to the metal plane. Under these constrains, we deduced the maximum possible $F_p$s ($F_p^{max}$s) by varying the position of the QD that is exactly resonant to the cavity mode. Figures 2(e) and (f) respectively summarize the calculated $F_p^{max}$s for the cavity modes based on the fundamental and higher-order SPP transverse mode. The estimated $F_p^{max}$s for the cavity designs with the fundamental SPP mode are not vanishing even for small waveguide width but are small (less than ~10) as expected from their electric field distributions. Meanwhile, $F_p^{max}$s for those with the higher-order SPP mode are much higher and reach ~24 when using a diameter of 2.0 μm and a waveguide width of 200 nm. These results demonstrate the advantage of the use of the higher-order SPP mode for defining plasmonic resonators with high Purcell effect for InAs/GaAs QDs with in-place optical transition dipoles. It is noteworthy that, for the designs with waveguide widths less than 180 nm, the higher-order SPP transverse mode is cut off. The cut-off width becomes 200 nm when considering microring structures with diameters less than 1.2 μm due to the sharp waveguide bending.

In order to investigate the optical properties of the fabricated samples, we conducted low temperature micro-PL measurements. The samples were placed in a liquid helium flow cryostat cooled to ~10 K and optically excited by a picosecond mode-locked laser (center wavelength = 780 nm, repetition rate ~ 80 MHz). The pump light was focused by an objective lens (numerical aperture = 0.65), and the PL emission from the sample was analyzed by a spectrometer equipped with an InGaAs array detector.

Firstly, we evaluated spectral mode distributions in the plasmonic resonators under relatively strong optical excitation with an average pump power of ~80 μW. We measured emission spectra for the plasmonic ring resonators with diameters from 1 to 4 μm while fixing the widths to be 300 nm. An example spectrum for the resonator with a diameter of 1 μm is shown in Figure 3(a). We observed multiple sharp peaks in the spectrum, which span nearly equidistantly and hence are assumed to primarily arise from the cavity modes based on the same transverse mode but with different azimuthal numbers. We compared the measured peak spacing, i.e. free spectral range (FSR, $\Delta\lambda$) with a standard model for microring resonators expressed as

$$\Delta\lambda = \frac{\lambda^2}{2\pi r}\left(n_{eff} - \lambda\frac{dn_{eff}}{d\lambda}\right)^{-1}, (2)$$

where r is the ring radius and $n_{eff}$ is the effective refractive index of the guided mode propagating in the SPP waveguide. Figure 3(b) summarizes the comparisons between the theoretical and the experimental FSRs for resonators with various diameters and a fixed width of 300 nm. The good agreement between the theory and experiments suggests the plasmonic origin of the observed emission peaks. We note that the $n_{eff}$s for the two SPP modes are very similar, thus hindering distinguishing the two modes solely by the FSRs.

Next, we performed time-resolved PL measurements of the QD radiation coupled to the cavity modes under the pulsed optical pumping with an average pump power on 1 μW. We employed a time correlated single photon counting apparatus composed of a superconducting single photon detector (SSPD). We studied a plasmonic resonator with a diameter of 1 μm and a width of 300 nm, in which only the two SPP transverse modes are supported and any other mode (including photonic modes) is cut off. Figure 3(c) shows a measured PL decay curve for the cavity emission at a wavelength of 1,067 nm with a Q factor of ~200. The decay curve was recorded with the help of a spectrometer-based bandpass filter at the cavity resonance wavelength with a bandwidth of ~0.05 nm. We fit the decay curve using an exponential decay function and extracted its lifetime to be ~0.35 ns. This is 3.7 times faster than that measured for QDs in an unstructured GaAs plate placed on the same silver surface, as plotted using blue dots in Figure 3(c). The reduced lifetime suggests the Purcell enhanced emission from the QDs in the plasmonic resonator.

We also investigated the wavelength dependence of QD's spontaneous emission rates for the same cavity mode by varying the center wavelength of the bandpass filtering. Figure 3(d) shows the evolution of PL decay rates around the cavity mode resonance. Overall, the decay rates peak around the cavity resonance wavelength, as expected for the Purcell-effect-enhanced emission processes. We compared the measured decay rates with a theoretical model based on eq 1 for the cavity modes respectively based on the fundamental and higher-order SPP mode, after taking into account their differences in field distribution and V. In Figure 3(d), we overlay the calculated emission decay rates for the cavity modes after averaging over all the possible QD positions in the QD layers. The decay rates have a constant offset of 1.18 ns$^{-1}$, which stems from the QD-to-free space coupling measured using the QDs in the unstructured GaAs plate on the silver. The theoretical curve for the higher-order SPP mode well describes the experimental results, allowing us for concluding that the observed enhancement of QD emission rate originates from the Purcell effect in the plasmonic cavity mode based on the higher-order SPP transverse mode.

Finally, we investigate single plasmon generation using the same plasmonic resonator with the diameter and width of 1 μm and 300 nm, respectively. The black curve in Figure 4(a) shows a PL spectrum of the target cavity mode taken under a strong average pump power of 80 μW, exhibiting a clear cavity resonance at 1067 nm. When reducing the pump power to 300 nW, we observed multiple sharp emission peaks originating from individual QDs, as can be seen in the blue curve in Figure 4(a). In the following, we focus on the bight emission peak at 1066.4 nm. First, we investigated its emission dynamics by measuring a time resolved PL spectrum, as plotted in Figure 4(b). Through fitting to the PL decay curve, we deduced its emission lifetime to be 0.46 ns, which indicates the accelerated emission process in the QD due to the Purcell effect with an $F_p$ of 2.7. This observation suggests that the dominant radiation from the QD directly couples to the plasmonic cavity mode and efficiently excites the SPPs in the plasmonic cavity. Then, we investigated the quantum nature of the generated SPPs in the resonator. We measured the intensity correlation between photons scattered out from the plasmonic mode. We employed a Hanbury Brown-Twiss setup equipped with two SSPDs. Figure 4(c) shows a recorded intensity correlation histogram. At the zero time delay, we observe a clear antibunching with a time origin value of the normalized second order coherence function of $g^{(2)}(0) = 0.27$. Combined with the measured $F_p$ of 2.7, these results demonstrate the Purcell-enhanced single plasmon generation in the plasmonic microring resonator.

In conclusion, we have demonstrated a SPG using a self-assembled InAs/GaAs QD embedded in a plasmonic microring resonator. The SPG was fabricated with the help of the transfer printing method, which enabled the facile three-dimensional stack of the GaAs microring on an atomically-flat metal surface. We theoretically and experimentally clarified that a plasmonic cavity mode based on the higher-order transverse SPP mode can support strong Purcell effect on the QD, due to its strong in-place electric field. The strong Purcell effect mediates direct coupling of the QD radiation into the SPP modes. Finally, we observed the single SPP generation in the plasmonic resonator by measuring a strong anti-bunching in an intensity correlation histogram. We believe that our approach is advantageous for pursuing deterministic, high-performance SPGs and for the integration of SPGs into scalable quantum plasmonic circuits. In this context, an important future work is to demonstrate indistinguishable single plasmon generation from QDs on the metalized chip, which constitutes a key test toward implementing large scale quantum plasmonic circuits.

FIGURES

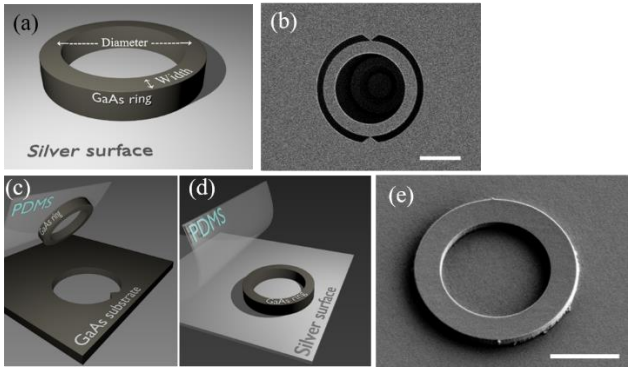

Figure 1. (a) Schematic image of the plasmonic microring resonator investigated in this study. The GaAs ring containing QDs is placed on the silver surface. (b) SEM images of the air-bridge microring resonator prepared for the subsequent transfer printing process. The scale bar corresponds to 500 nm. Schematic illustration of (c) the pick-up process of a microring using a stamp glue (PDMS) and of (d) the lift-off process of the microring onto the silver surface. (e) SEM image of a transfer-printed microring on the silver surface. The scale bar corresponds to 500 nm.

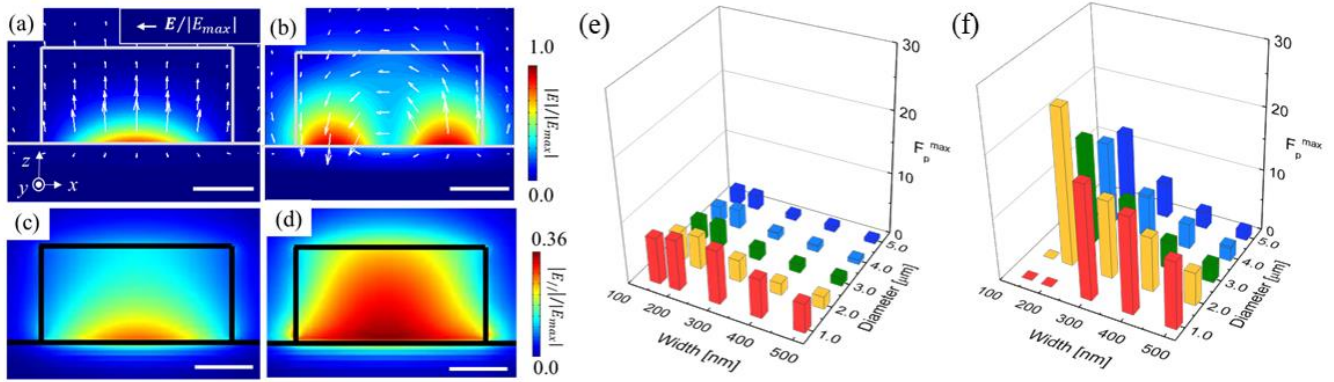

Figure 2. Cross-sectional electric field distributions obtained by the waveguide mode simulations for the (a) fundamental and (b) higher-order transverse mode. The waveguide height is 150 nm and the width is 300 nm. The white arrows exhibit the intensities and directions of the electric field vectors for each mode, which are constructed from the electric fields lying within the cross section ($E_x$ and $E_z$). (c) and (d) Simulated in-plain electric field distributions for the fundamental and higher-order transverse modes, respectively. The electric field components directed to the wave propagation direction ($E_y$) are considered in the plots. Scale bars in (a)-(d) indicate 100 nm. (e) Calculated maximum possible Purcell factors ($F_p^{max}$s) for the embedded QDs in the resonators based on the fundamental SPP mode, designed with various sets of ring diameters and waveguide widths. (f) The same in (e) but for those based on the higher-order SPP mode.

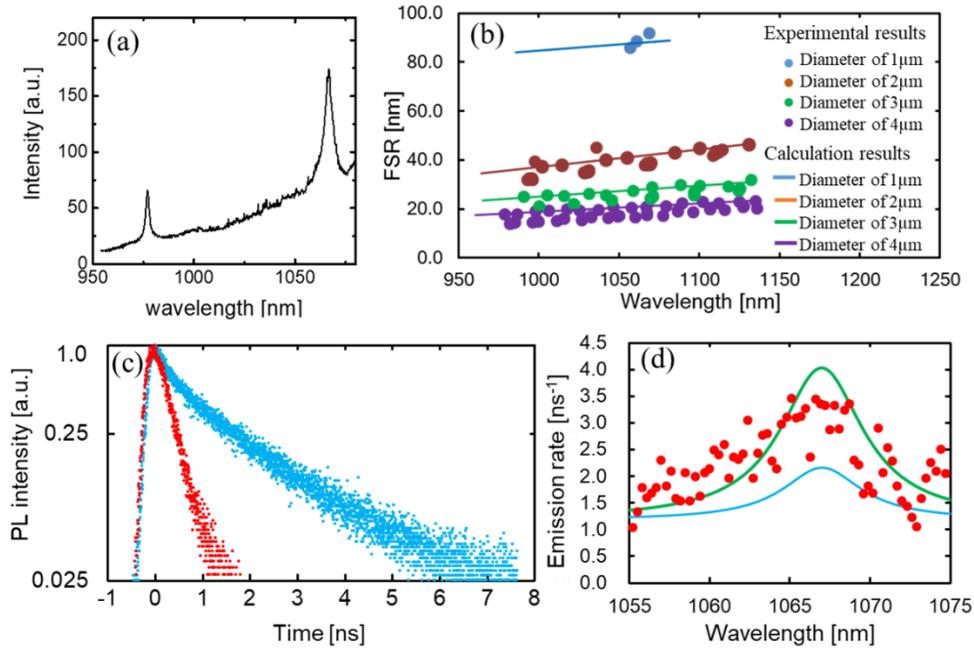

Figure 3. (a) Emission spectrum under a strong excitation condition with an average pump power of 80 μW. The center wavelengths of the observed peaks are 977 nm and 1067 nm. (b) Measured FSRs (colored points) plotted as a function of emission peak wavelengths. Solid lines exhibit simulated FSRs for the corresponding plasmonic cavity modes. (c) Measured PL decay curves. The red points show results for the plasmonic cavity mode resonating at 1067 nm. The blue points show a decay curve recorded for a QD embedded within an unstructured GaAs plate placed on the same silver surface. (d) Summary of measured PL emission rates around the plasmonic cavity resonance at 1067 nm (red points). The colored solid lines show the calculated PL decay rates for the cavity modes respectively based on the fundamental (blue) and the higher-order (green) transverse mode.

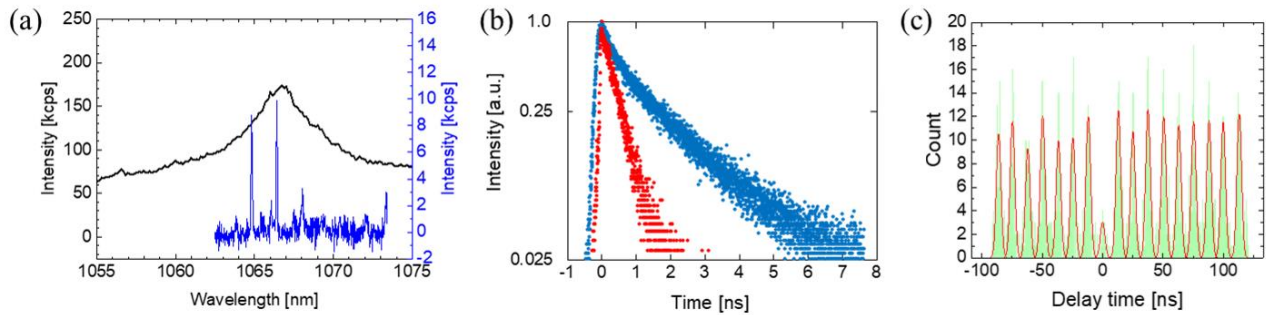

Figure 4. (a) Emission spectra of the plasmonic cavity. The black curve shows a spectrum under a strong pumping condition with an average pump power of 80 μW. The blue curve shows that under a weak excitation with 300 nW pumping, exhibiting narrow emission peaks stemming from presumably-individual QDs. The emission wavelength of the single QD is 1066.4 nm, which is close to the cavity resonance wavelength of ~1067 nm. (b) Measured time-resolved PL spectra. The red curve shows the PL decay curve for the target QD coupled to the cavity mode. The blue curve is of the QD embedded in an unstructured GaAs slab on the silver surface. (c) Second order intensity correlation histogram measured for the signal from the target QD. We indirectly measured the quantum property of the SPPs generated from the QD via scattered photons out of the plasmonic cavity mode. The red line shows the fitting curve. The time origin value of the second order coherence function, $g^{(2)}(0)$, is deduced to be 0.266, demonstrating single SPP generation in the resonator.

## ACKNOWLEDGMENT

The authors thank R. Katsumi for technical supports on transfer printing. This work was supported by the Japan Society for the Promotion of Science (JSPS) KAKENHI Grant-in-Aid for Specially Promoted Research (15H05700), KAKENHI (16K06294) and JST PRESTO (JPMJPR1863), and is based on results obtained from a project commissioned by the New Energy and Industrial Technology Development Organization (NEDO).